\documentstyle[12pt,epsf,epsfig]{article}
\newcount\timecount
\newcount\hours \newcount\minutes  \newcount\temp \newcount\pmhours

\hours = \time
\divide\hours by 60
\temp = \hours
\multiply\temp by 60
\minutes = \time
\advance\minutes by -\temp
\def\hour{\the\hours}
\def\minute{\ifnum\minutes<10 0\the\minutes
            \else\the\minutes\fi}
\def\clock{
\ifnum\hours=0 12:\minute\ AM
\else\ifnum\hours<12 \hour:\minute\ AM
      \else\ifnum\hours=12 12:\minute\ PM
            \else\ifnum\hours>12
                 \pmhours=\hours
                 \advance\pmhours by -12
                 \the\pmhours:\minute\ PM
                 \fi
            \fi
      \fi
\fi
}

\def\monthname{\relax\ifcase\month 0/\or January\or February\or
   March\or April\or May\or June\or July\or August\or September\or
   October\or November\or December\else\number\month/\fi}

\def\bold#1{\setbox0=\hbox{$#1$}%
     \kern-.025em\copy0\kern-\wd0
     \kern.05em\copy0\kern-\wd0
     \kern-.025em\raise.0433em\box0 }

\def\gappeq{\mathrel{\rlap {\raise.5ex\hbox{$>$}}
{\lower.5ex\hbox{$\sim$}}}}

\def\lappeq{\mathrel{\rlap{\raise.5ex\hbox{$<$}}
{\lower.5ex\hbox{$\sim$}}}}

\def\ga{\mathrel{\raise.3ex\hbox{$>$\kern-.75em\lower1ex\hbox{$\sim$}}}}
\def\la{\mathrel{\raise.3ex\hbox{$<$\kern-.75em\lower1ex\hbox{$\sim$}}}}
\def\gev{{\rm \, Ge\kern-0.125em V}}
\def\tev{{\rm \, Te\kern-0.125em V}}
\def\beq{\begin{equation}}
\def\eeq{\end{equation}}

\def\m12{m_{1\!/2}}

\begin{document}
\begin{titlepage}
\pagestyle{empty}
\baselineskip=21pt
\rightline{hep-ph/0109288}
\rightline{CERN--TH/2001-256}
\rightline{ACT-09/01, CTP-TAMU-29/01}
\rightline{UMN--TH--2025/01, TPI--MINN--01/43}
\vskip 0.05in
\begin{center}
{\large{\bf Lower Limits on Soft Supersymmetry-Breaking Scalar Masses}}
\end{center}
\begin{center}
\vskip 0.05in
{{\bf John Ellis}$^1$, 
{\bf D.V. Nanopoulos}$^2$ and
{\bf Keith A.~Olive}$^{3}$
\vskip 0.05in
{\it
$^1${TH Division, CERN, Geneva, Switzerland}\\
$^2${Department of Physics, Texas A \& M University,
College Station, TX~77843, USA; \\
Astroparticle Physics Group, Houston
Advanced Research Center (HARC), \\
Mitchell Campus,
Woodlands, TX~77381, USA; \\
Chair of Theoretical Physics,
Academy of Athens,
Division of Natural Sciences,  
28~Panepistimiou Avenue,
Athens 10679, Greece}\\
$^3${Theoretical Physics Institute, School of Physics and Astronomy,\\
University of Minnesota, Minneapolis, MN 55455, USA}\\
}}
\vskip 0.05in
{\bf Abstract}
\end{center}
\baselineskip=18pt \noindent

Working in the context of the CMSSM, we argue that phenomenological
constraints now require the universal soft supersymmetry-breaking scalar
mass $m_0$ be non-zero at the input GUT scale. This conclusion is
primarily imposed by the LEP lower limit on the Higgs mass and the
requirement that the lightest supersymmetric particle not be charged. We
find that $m_0 > 0$ for all $\tan \beta$ if $\mu < 0$, and $m_0 = 0$ may
be allowed for
$\mu > 0$ only when $\tan \beta \sim 8$ and one allows an uncertainty of
3+ GeV in the theoretical calculation of the Higgs mass. Upper limits on
flavour-changing neutral interactions in the MSSM squark sector allow
substantial violations of non-universality in the $m_0$ values, even if
their magnitudes are comparable to the lower limit we find in the CMSSM.  
Also, we show that our lower limit on $m_0$ at the GUT scale in the CMSSM
is compatible with the no-scale boundary condition $m_0 = 0$ at the Planck
scale.

\vfill
\vskip 0.15in
\leftline{CERN--TH/2001-256}
\leftline{August 2001}
\end{titlepage}
\baselineskip=18pt

Motivated by the naturalness of the gauge hierarchy~\cite{hierarchy},
TeV-scale supersymmetry is, perhaps, the most plausible scenario for
low-energy physics beyond the Standard Model. Here we study the minimal
supersymmetric extension of the Standard model (MSSM). Some of the
greatest puzzles of supersymmetry are associated with its breaking. There
is no consensus on the origin of supersymmetry breaking, even within
string (or M) theory, and we do not know what fixes the scale of
supersymmetry breaking (and how). Within this general area of puzzles,
there are minor puzzles, such as questions whether soft
supersymmetry-breaking scalar and gaugino masses, $m_0$ and $m_{1/2}$,
respectively, are universal. In particular, generation-dependent scalar
masses would threaten the observed suppression of flavour-changing neutral
interactions (FCNI)~\cite{EN}, whereas differences between the scalar
masses of sparticles with different gauge quantum numbers would be less
problematic. In other words, one needs $m_0^{e_{L}} = m_0^{\mu_{L}} =
m_0^{\tau_{L}}$ to a very good approximation, and similarly for the
$\ell_R$ and for the squarks. On the other hand, there is no strong
phenomenological reason why $m_0^{\ell_{L}} = m_0^{\ell_{R}}$, or why
squark and slepton masses should be equal. For the moment, however, we
work in the context of the constrained MSSM (CMSSM), where this extended
universality is assumed.

Some proposed mechanisms for supersymmetry breaking in string theory yield
generation-dependent scalar masses, for example because they depend on
moduli characterizing the string vacuum, whereas other mechanisms are
naturally generation-independent. The former are {\it a priori} in
conflict with the constraints imposed by FCNI. Many of the latter
mechanisms achieve consistency with these limits by resuscitating no-scale
gravity~\cite{noscale}, in which the soft supersymmetry-breaking scalar
masses {\it vanish at the input supersymmetric grand-unification (GUT)
scale}. These input values are renormalized by gauge and Yukawa
interactions at lower scales. The renormalizations by gauge interactions
are generation-independent, whereas those by Yukawa interactions break
universality by amounts related to quark and lepton masses, which may be
phenomenologically acceptable. Therefore, no-scale supergravity and its
possible string antecedents are experiencing some sort of phenomenological
renaissance.

Is the no-scale hypothesis italicized in the previous paragraph actually
excluded by the continuing absence of sparticles and by other experimental
limits? Or does it require some reformulation? If so, are the FCNI 
constraints endangered? These are the issues addressed in this paper.

As we have indicated, we
restrict our attention to the CMSSM which imposes
universal gaugino masses $m_{1/2}$,
scalar masses $m_0$ (including those of the Higgs multiplets) and
trilinear supersymmetry breaking parameters $A_0$ are input at the
supersymmetric grand unification scale. In this framework, the Higgs
mixing parameter $\mu$ can be derived (up to a sign) from the other MSSM
parameters by imposing the electroweak vacuum conditions for any given
value of $\tan \beta$.  Thus, given the set of input parameters determined
by $\{ m_{1/2}, m_0, A_0,\tan\beta,sgn(\mu) \}$, the entire spectrum of
sparticles can be derived.  
In our analysis, we consider the following experimental limits. (1) Lower
limits on slepton masses from LEP, in particular the bound $m_{\tilde
e_R} > 99$~GeV~\cite{LEPslepton}. (2)  The LEP lower limit on mass of the
lightest Higgs boson: $m_h > 114.1$~GeV~\cite{LEPH} - where we discuss and
take into account theoretical uncertainties in the mass
calculation~\cite{FeynHiggs}. (3)  The experimental range for $b \to s
\gamma$ decay~\cite{bsgamma}. (4) The recent BNL E821 measurement of the
anomalous magnetic moment of the muon~\cite{g-2} - however, we are
reluctant to rely strongly on this latter constraint until the theoretical
uncertainties in the Standard Model prediction are more widely understood.
(5) The requirement that the lightest supersymmetric particle (LSP) be the
lightest neutralino $\chi$~\cite{EHNOS}, rather than the lightest slepton
$\tilde \tau_1$. (6) The lower limit $\Omega_\chi h^2 > 0.1$ on the relic
LSP density, which would apply if the LSP constitutes most of the cold
dark matter in the Universe, though we allow this condition to be
relaxed.

Since the last two constraints are less directly related to particle
experiments, we motivate them in more detail. Recent observations in
cosmology and particle physics strengthen the expectation that
supersymmetry plays a fundamental r\^{o}le in the structure of the
Universe. The observation of the first three acoustic peaks in the Cosmic
Microwave Background (CMB) radiation anisotropies~\cite{CMB} not only
indicates clearly that we are living in (or very near) a flat $k=0$ or
$\Omega$ =1 Universe, as indicated by the position of the first peak, but
also confirms that the mass-energy density of the Universe is mostly
non-baryonic, as indicated by the ratios of the heights of the even and
odd peaks. These indicate that the baryon density $\Omega_b h^2 \sim
0.021$, with an uncertainty of 10 to 20\%, and is in good agreement with
the entirely independent estimates from Big-Bang
nucleosynthesis~\cite{CFO} based on the abundance of D/H in quasar
absorption systems \cite{omear}. The combination of CMB measurements
together with other astrophysical data independently require a much
larger total for the matter density,
$\Omega_m \sim 0.3$, also in agreement with previous independent
estimates. The case for non-baryonic dark matter has therefore been
greatly strengthened by these recent measurements of the CMB.

It is well known that low-energy supersymmetry: $m_{susy} = {\cal
O}(1)$~TeV, as motivated independently by the gauge hierarchy
problem~\cite{hierarchy} and the unification of gauge couplings
\cite{corner}, provides an excellent candidate for this non-baryonic dark
matter, namely the LSP~\cite{EHNOS}, as long as $R$ parity is conserved,
as we assume in this paper~\footnote{FCNI pose even more challenges for
$R$-violating theories.}. However, this LSP cannot be charged, or it
would conflict strongly with upper limits on charged relics from the Big
Bang. This motivates requirement (5) above,
$\tilde \tau_1 > m_\chi$. The lightest neutralino has all the properties
desired for non-baryonic cold dark matter, which should have
$\Omega_{CDM} h^2 > 0.1$, although there could be other components, in
which case $\Omega_\chi h^2 < 0.1$ might be possible. If the LSP is the
dominant component of the cold dark matter, the allowed range for its
relic density: $0.1 \leq
\Omega_{\chi} h^2 \leq 0.3$ provides stringent constraints on the
sparticle masses and thus on the soft supersymmetry-breaking 
parameters. In particular, neutralino annihilation depends on the sfermion
masses, and hence on the soft scalar masses $m_0$. Hence, the lower bound
$\Omega_{\chi}h^2 > 0.1$ could be translated into an upper bound on the
$\chi - \chi$ annihilation cross section, which in turn would imply a
lower bound on $m_0$, as mentioned in point (6) above. In fact, as we show
below, there is a non-trivial lower bound on $m_0$ in the CMSSM, even if
one only imposes the weaker requirement $\tilde \tau_1 > m_\chi$.

\begin{figure}[htb]
\begin{center}
\mbox{\hskip -.2in \epsfig{file=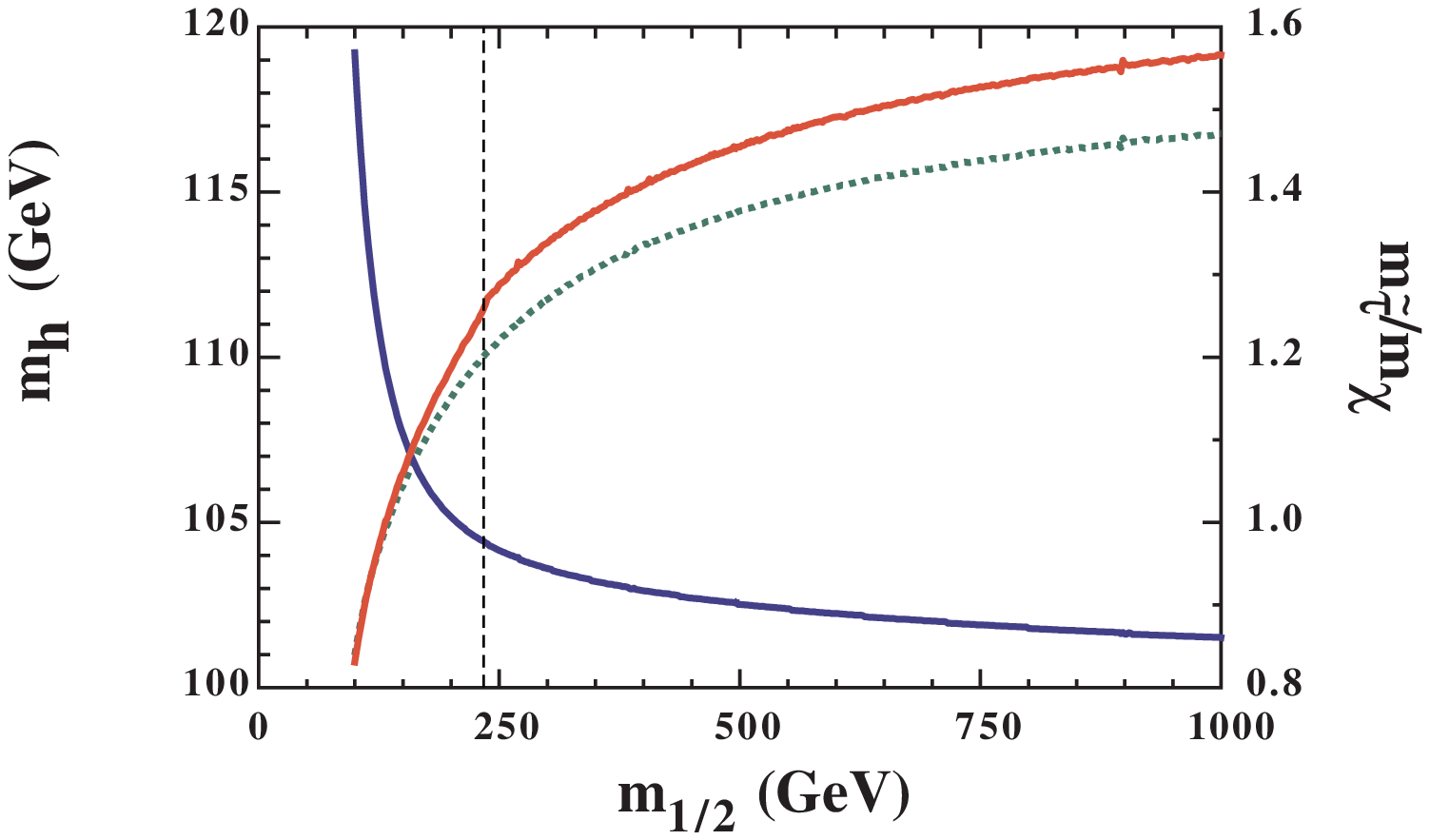,height=9cm}}
\mbox{\hskip -.2in \epsfig{file=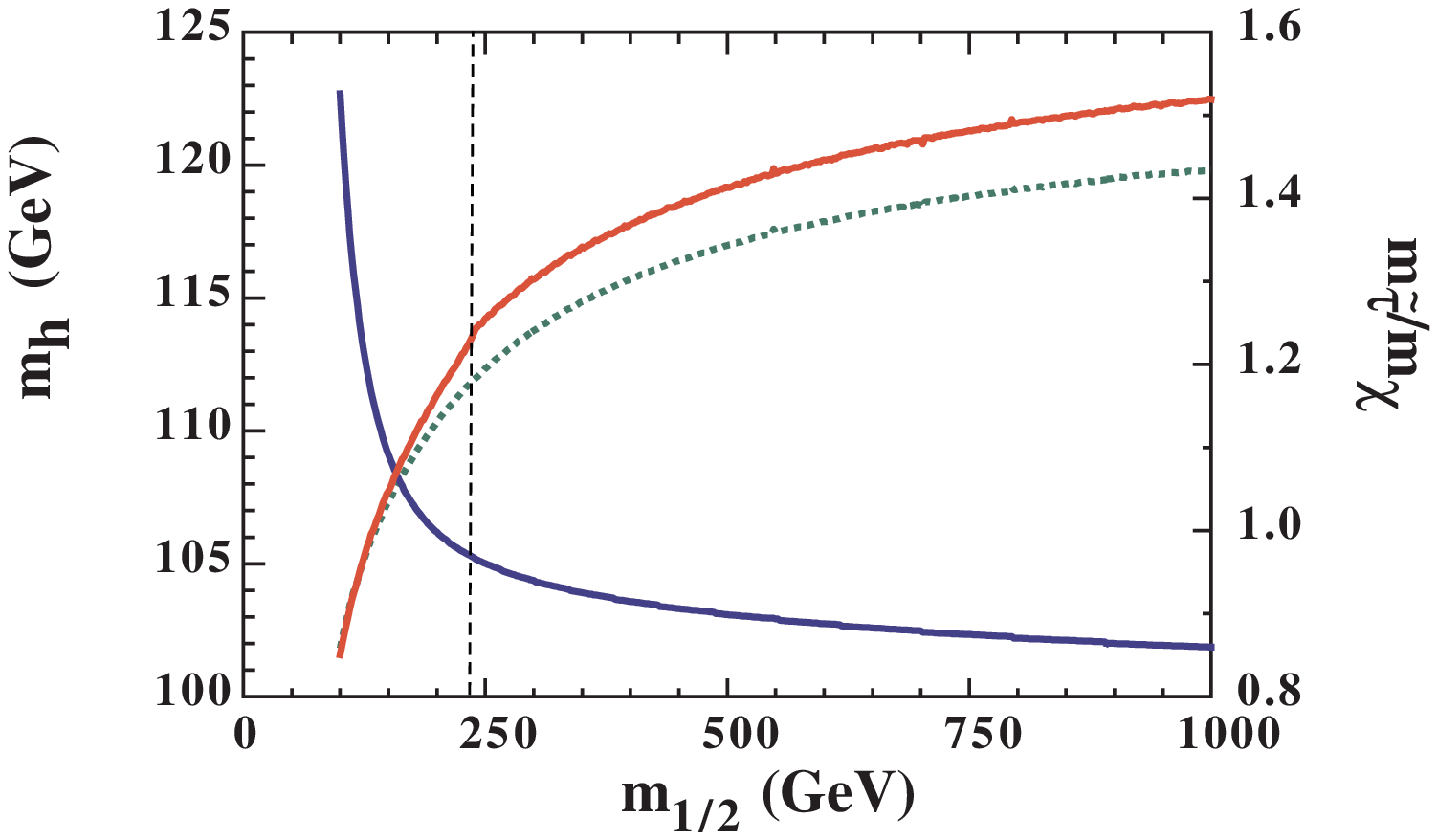,height=9cm}}
\end{center}
\caption[.]{
\label{fig:m0tb10}
{\it The price of requiring $m_0 = 0$, showing the interplay of the 
constraints (i) to (v) for (a) $\tan \beta =
10, \mu > 0$ and $m_t = 175$~GeV, and (b) for $\tan \beta =
10, \mu > 0$ and $m_t = 180$~GeV. The red solid (green dashed) curves with
positive slope show the calculated value of $m_h$ using the
{\tt FeynHiggs} code \cite{FeynHiggs} (using HHH \cite{HHH}). The blue
solid curve with negative slope shows the ratio $m_{\tilde \tau}/m_\chi$.
The vertical thin dashed line shows the lower limit on $m_{1/2}$ due to
the selectron mass limit. }}
\end{figure}

A first example of the interplay between these constraints is shown in
Fig.~\ref{fig:m0tb10}(a). Here $m_0 = 0$ is assumed, as are $\tan \beta =
10, \mu > 0$ and $m_t = 175$~GeV. The left-hand vertical axis shows the
value of $m_h$, and the right-hand side shows the ratio $m_{\tilde \tau_1}
/ m_\chi$. We indicate the impact of constraint (1)
above, namely that $m_{e_R} > 99$~GeV~\cite{LEPslepton} by the vertical
thin dashed line. For $m_0 = 0$ and $\tan \beta = 10$, the selectron
mass limit implies $m_{1/2} > 230$ GeV.  The impact of constraint (2)
above depends on the codes used to evaluate
$m_h$. We show as a solid (red) line the value calculated with the {\tt
FeynHiggs} code~\cite{FeynHiggs}, and as a dashed (green) line the value
calculated with the program of~\cite{HHH}, hereafter referred to as {\tt
HHH}. They differ little for $m_{1/2} < 200$~GeV and/or $m_h < 110$~GeV,
but disagree by up to $\sim 3$~GeV at large $m_{1/2}$. We find that the
experimental limit $m_h > 114.1$~GeV imposes $m_{1/2} > 330 (465)$~GeV if
we use the {\tt FeynHiggs} ({\tt HHH}) code. There is no significant
constraint (3) from $b \to s \gamma$ for this value of $\tan \beta = 10$
and sign of
$\mu$. The measured value (4) of $g_\mu - 2$ favours the range $175~{\rm
GeV} < m_{1/2} < 450$~GeV ($195~{\rm GeV} < m_{1/2} < 290$~GeV) at the
two- (one-) $\sigma$ level. The constraint (5) $m_{\tilde \tau_1} / m_\chi
> 1$ imposes $m_{1/2} < 210$~GeV. Combining all these calculations, we
find {\it no range} of $m_{1/2}$ for which all these constraints are
satisfied. Specifically, the upper limit (5) from $m_{\tilde \tau_1} /
m_\chi$: $m_{1/2} < 210$~GeV is in {\it prima facie} contradiction with
the lower limit (2) from $m_h$: $m_{1/2} > 330 (465)$~GeV. Phrased another
way: $m_{\tilde \tau_1} / m_\chi > 1$ only for values of $m_{1/2}$
corresponding to $m_h < 110$~GeV, which is excluded for the CMSSM
discussed here.

Since calculations of $m_h$ are well known to be very sensitive to $m_t$,
with $\partial m_h / \partial m_t > 0$, we show in
Fig.~\ref{fig:m0tb10}(b) the equivalent of Fig.~\ref{fig:m0tb10}(a) for
$m_t = 180$~GeV. We see that the lower limit on $m_{1/2}$ becomes $250
(315)$~GeV for $m_h > 114.1$~GeV (2), whereas the other constraints are
essentially unchanged. Phrased another way: the lower limit on $m_h$
imposes $m_{\tilde \tau_1} / m_\chi < 0.94$, even in the more conservative
{\tt FeynHiggs} calculation~\cite{FeynHiggs}. {\it We conclude that $m_0 =
0$ is not possible for $\tan \beta = 10$ and $\mu > 0$.}

\begin{figure}
\vspace*{-0.75in}
\hspace*{-.10in}
\begin{minipage}{8in}
\epsfig{file=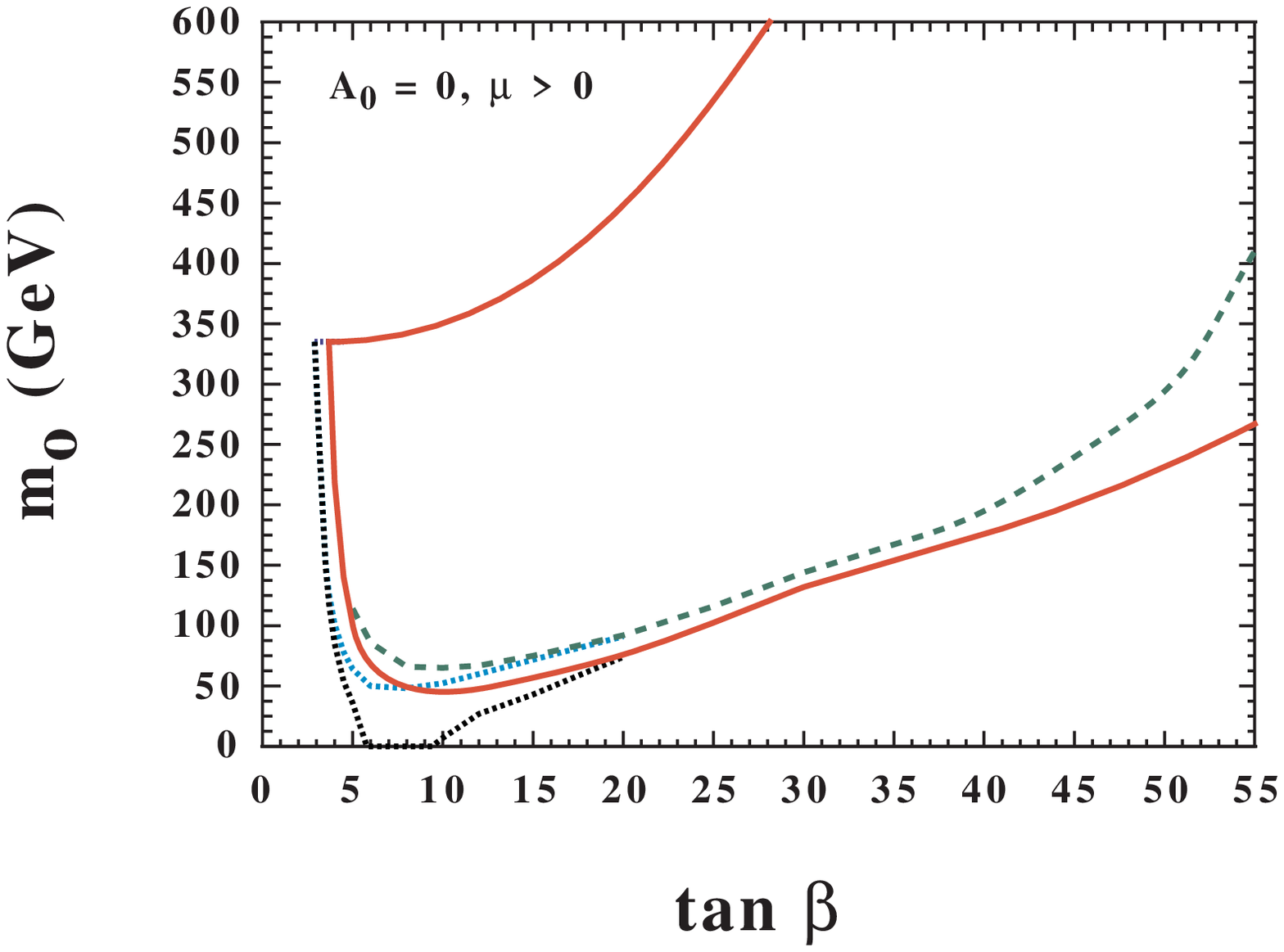,height=2.5in}
\epsfig{file=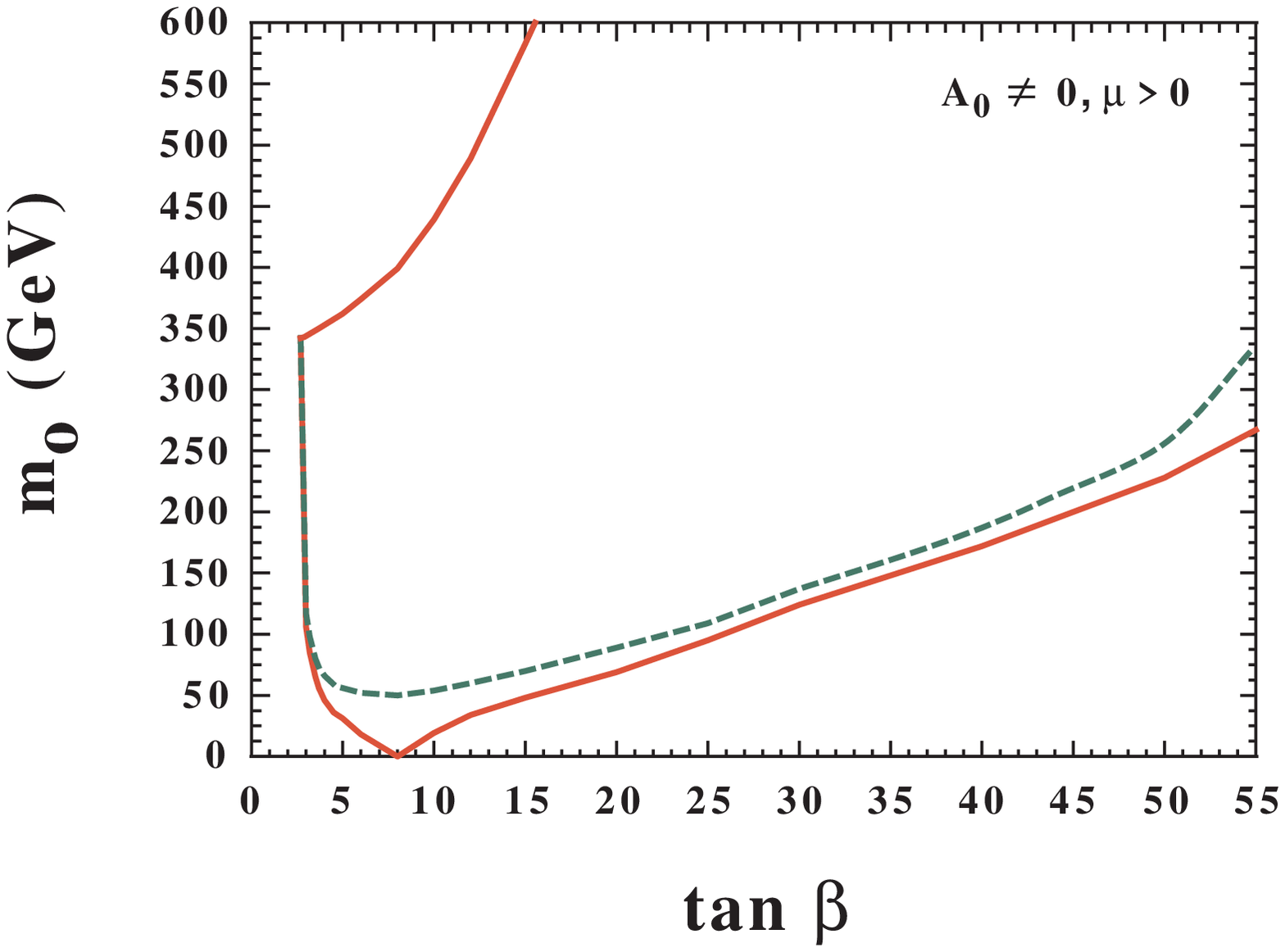,height=2.5in} \hfill
\end{minipage}
\begin{minipage}{8in}
\epsfig{file=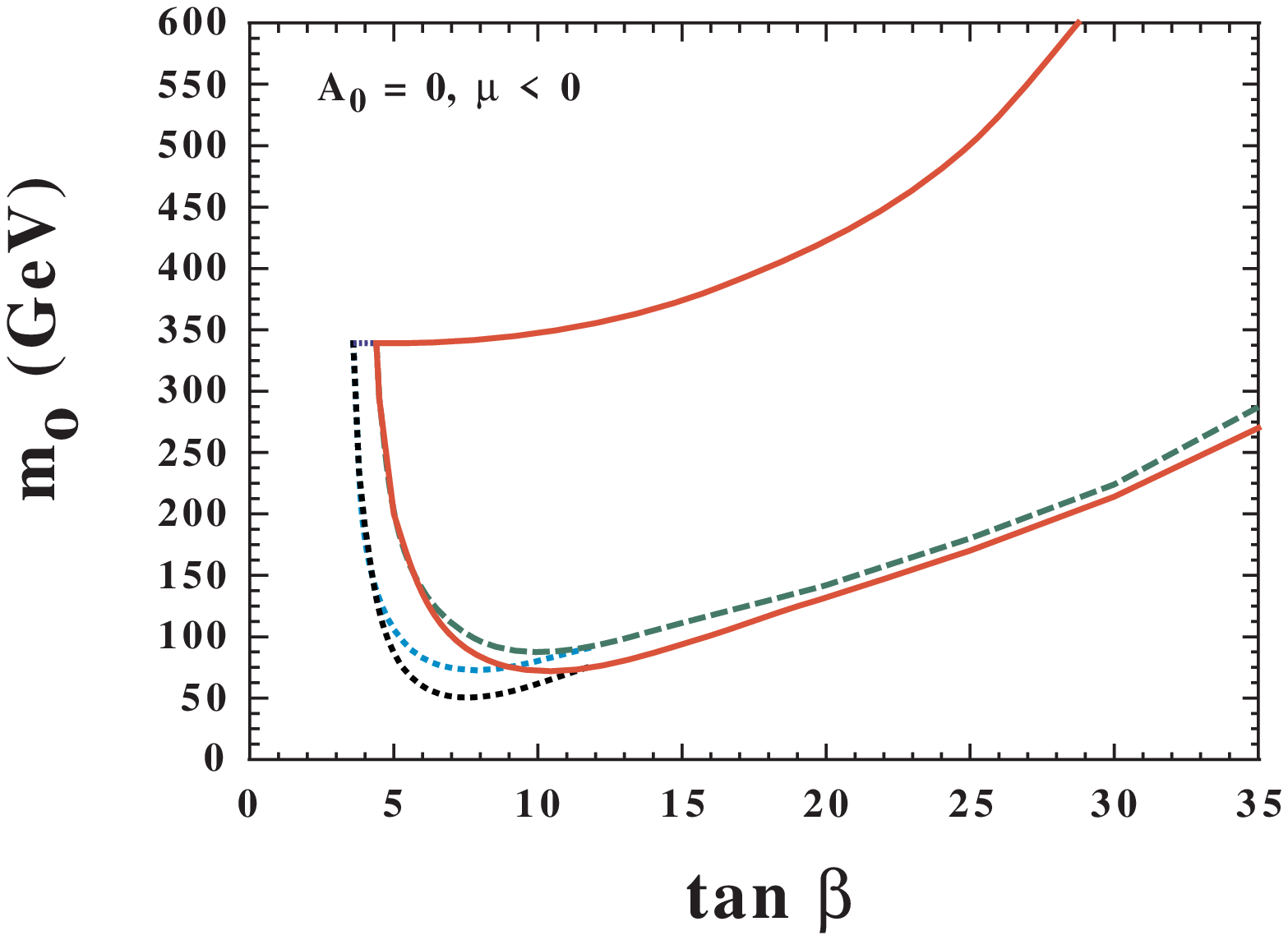,height=2.5in}
\epsfig{file=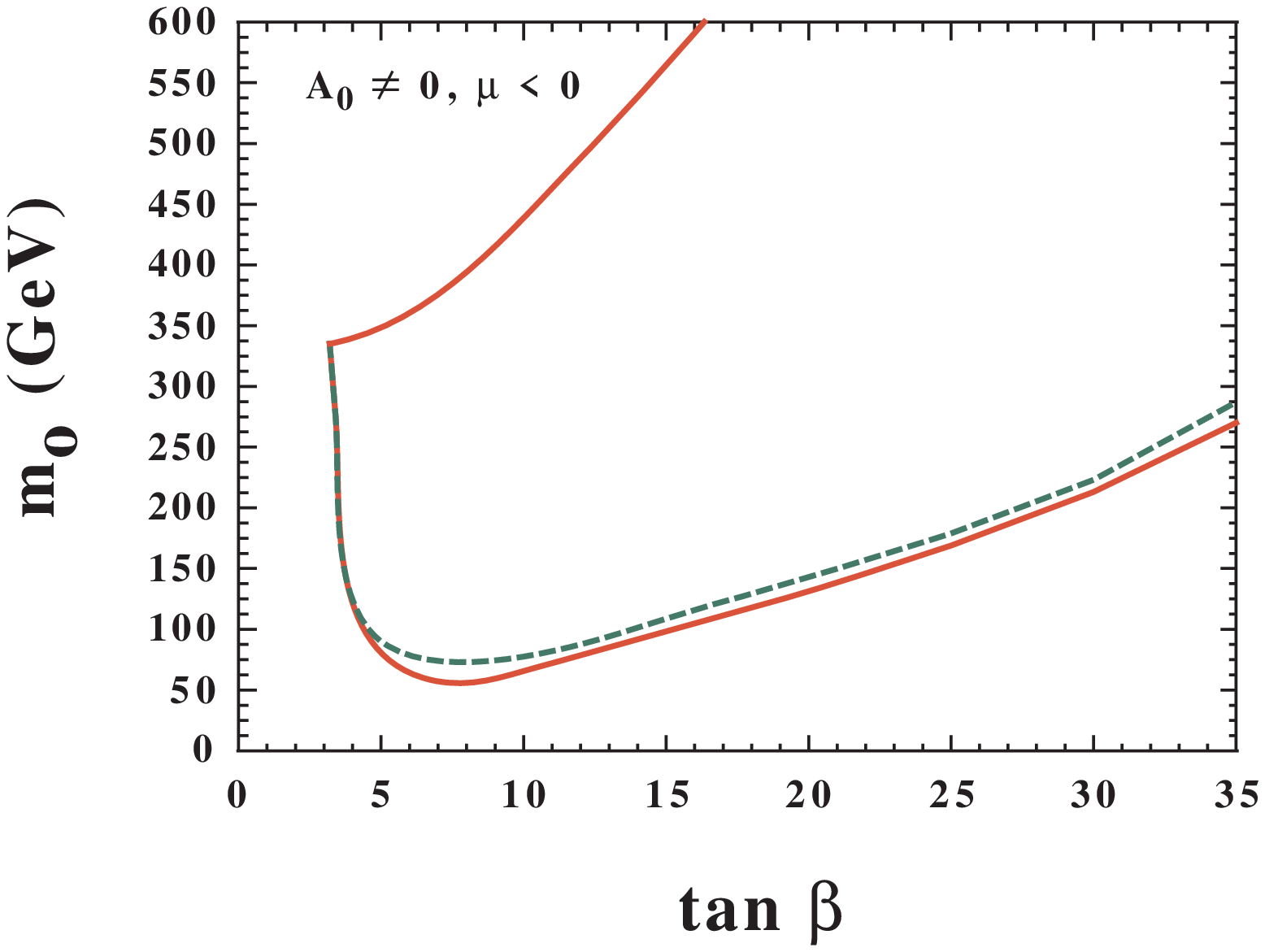,height=2.5in} \hfill
\end{minipage}
\caption{\label{fig:lowerm0}
{\it Lower and upper limits on $m_0$ due to the different constraints, for 
(a) $\mu > 0$ and $A_0 = 0$, (b) $\mu > 0$ and $A_0 \ne 0$, 
(c) $\mu < 0$ and $A_0 = 0$, (b) $\mu < 0$ and $A_0 \ne 0$. The solid 
(red) line and dashed (green) are for $m_h > 113.5$~GeV, the dotted
lines for $m_h > 110.5$~GeV, as calculated using the {\tt
FeynHiggs}  code~\cite{FeynHiggs}.
The solid (red) and darker dotted (black) lines only require $m_{\tilde 
\tau_1} > m_\chi$, whereas the dashed (green) and lighter dotted (blue)
lines require 
$\Omega_\chi h^2 > 0.1$.
}}
\end{figure}

More general views of the interplays between the different constraints (1)  
to (6) above as functions of $\tan \beta$ are shown in
Fig.~\ref{fig:lowerm0}. Panel (a) is for $\mu > 0$ and $A_0 = 0$, with
$m_t = 175$~GeV. We restrict the analysis to $\tan \beta \le 55$, which is 
close to the largest value for which we find generic regions of parameter
space  with consistent electroweak symmetry breaking~\cite{EFGOSi}. Four
lower  limits on $m_0$ 
are plotted, corresponding to different implementations of the 
constraints. The (red) solid line assumes $m_h > 
113.5$~GeV~\footnote{Calculated 
(conservatively) with the {\tt FeynHiggs} code~\cite{FeynHiggs}.}, which 
allows 
some safety factor compared with the experimental lower limit of 
114.1~GeV, 
and employs the weaker cosmological constraint (5) $m_{\tilde \tau_1} > 
m_\chi$ (i.e., it ignores the constraint (6)). The (green) dashed line
also uses
$m_h > 113.5$~GeV, but imposes  the stronger cosmological constraint (6)
$\Omega_\chi h^2 > 0.1$. We see  in both cases that $m_0 \ne 0$: the
absolute lower limits are
\begin{eqnarray}
m_0 \; & \ga & \; 40~{\rm GeV~for}~m_{\tilde \tau_1} > m_\chi,  
\\
m_0 \; & \ga & \; 65~{\rm GeV~for}~\Omega_\chi h^2 > 0.1,
\label{lowermzero}
\end{eqnarray}
both attained for $\tan \beta \sim 8$ to 10. The rise in the lower bound 
on $m_0$ for smaller $\tan \beta$ reflects the impact of the $m_h$ 
constraint. This constraint is also important for $\tan \beta \la 15$ to 
20, but the lower limit for larger $\tan \beta$ reflects the impact of the 
$b \to s \gamma$ constraint combined with the weaker or stronger 
cosmological constraint. The darker (black) dotted line employs 
$m_{\tilde \tau_1} > m_\chi$ and the weaker 
constraint $m_h > 110.5$~GeV, which allows a generous safety 
theoretical factor compared with the available codes {\tt FeynHiggs} and 
{\tt HHH}. We recall - see Fig.~\ref{fig:m0tb10} - that these codes agree 
rather well for small $m_0$ and $m_{1/2}$ and $\tan \beta \sim 10$. This 
is the only case where $m_0 = 0$ may be permitted, and only for $6 \la
\tan 
\beta \la 9$. If one strengthens the cosmological constraint to require 
$\Omega_\chi h^2 > 0.1$, $m_0 = 0$ is disallowed even for the weak Higgs 
constraint, as shown by the lighter (blue) dotted curve. The convergence 
of the curves corresponding to the weaker and stronger Higgs constraints 
for $\tan \beta \ga $  20 reflects the dominance of the $b \to s 
\gamma$ constraint at larger $\tan \beta$.

Panel (b) of Fig.~\ref{fig:lowerm0} is obtained by varying $A_0 \ne 0$.
When looking for the minimum of $m_0$, $m_t = 180$~GeV was chosen, as that
always weakens the Higgs mass constraint on $m_{1/2}$, and $m_0$ is then
generally allowed to be smaller, as seen in Fig.~\ref{fig:m0tb10}. Then
$A_0$ was varied so as to minimize the allowed value of $m_0$. When $\tan
\beta$ is small, $m_{1/2}$ is nevertheless forced to be high, approaching
the end of the coannihilation region \cite{efosi}. In this case, the
minimum and maximum values of $m_0$ become the same.  As $\tan \beta$ is
raised,
$m_{1/2}$ continues to be constrained by the lower limit on $m_h$, but the
lower limit generally decreases. Increasing $A_0$ enhances this
tendency, but $A_0$ cannot be made arbitrarily large, for fear of driving
some scalars tachyonic. As $\tan\beta$ increases and $m_{1/2}$ is lowered,
the value of $A_0$ used also drops. At intermediate values of $\tan
\beta$, the effect of $A_0$ on $m_{\tilde \tau_1}$ also becomes relevant.
Increasing $A_0$ tends to reduce $m_{\tilde \tau_1}$, which must be
compensated by raising $m_0$. Hence, there is competition between wanting
$A_0$ large so as to increase $m_h$ and small so as to increase $m_{\tilde
\tau_1}$. For $\tan \beta \sim 8$, the minimum values of $m_0$ are found
when $A_0 \sim 0$, whereas $A_0 < 0$ is preferred for larger $\tan \beta$.
The $b \to s \gamma$ constraint also becomes relevant for larger $\tan
\beta$, and the minimum value of $m_0$ is generally found when $A_0 \sim
- 150$~GeV. In Fig.~\ref{fig:lowerm0}b, we show only the stronger
Higgs mass bound, $m_h > 113.5$ GeV.  The solid (red) curves ignores the
cosmological constraint (6), whereas the dashed (green) curve includes it.
We see that when the constraint (6) is ignored, there is a small range in
$\tan \beta \sim 8$ where $m_0 = 0$ is allowed.

Also shown in Fig.~\ref{fig:lowerm0} are upper limits on $m_0$. These
apply to the `bulk', `coannihilation' and rapid-annihilation `funnel'
regions allowed by cosmology~\cite{EFGOSi}, but not to the `focus-point'
region~\cite{FM}. The latter is typically a narrow strip in the $(m_{1/2},
m_0)$ plane that appears at much larger $m_0$ than the range studied here,
the precise location being quite uncertain, being rather sensitive to the
choices of input parameters, particularly the top quark mass, and
higher-order effects in the model~\cite{EO}. Setting aside the
focus-point region, the maximum value of $m_0$ is always given by the tip
of the coannihilation region.  The mass of the lighter stau increases
with $A_0$, and the position of the coannihilation tail also scales with
$A_0$. The maximum value we have considered is $A_0 = 3$~TeV. This choice
is somewhat arbitrary, but seems to us relatively conservative.

Panels (c) and (d) of Fig.~\ref{fig:lowerm0} are for $\mu < 0$, with $A_0
= 0$ and $A_0 \ne 0$, respectively. They show similar qualitative features
to the corresponding panels (a) and (b) for $\mu > 0$, with the notable
exception that $m_0 = 0$ is disallowed, even if one uses only the weaker
cosmological bound $m_{\tilde \tau_1} > m_\chi$.
Because of the correlation between $\mu$ and the anomalous magnetic
moment of the muon, constraint (4) is not satisfied for $\mu < 0$.
Note also the increased importance of the $b \to s \gamma$ constraint
is seen by the merging of the curves in panel (c) at lower $\tan \beta$ 
than in (a). 

We also see in Fig.~\ref{fig:lowerm0} a lower bound on $\tan \beta$, which
is $\sim 3$ for $\mu > 0$ and the weaker Higgs mass bound $m_h > 110.5$
GeV, rising to
$\sim 4$ for the stronger requirement $m_h > 113.5$ GeV. The lower limit
would be much stronger, $\tan \beta \sim 8$, if one required constraint
(4) coming from the anomalous magnetic moment of the muon.

There are several issues that can be addressed in view of the allowed
region for $m_0$ seen in Fig.~\ref{fig:lowerm0}. Reaches for the discovery
of sparticles has been discussed rather extensively recently~\cite{Bench}, 
so we
concentrate our attention on other issues. One is the menace of 
FCNI~\cite{EN} that
was mentioned earlier in this paper. One may ask whether the constraints
discussed above allow $m_0 / m_{1/2}$ to be sufficiently small for (at
least some of) the FCNI constraints to be obeyed in a natural way. 
Consider, for example, the constraint imposed by the real part of
$K^0 - \overline K^0$ mixing~\footnote{We do not consider the imaginary 
part of $K^0 - \overline K^0$ mixing, regarding CP violation in the 
MSSM as an independent challenge.}. One should consider box diagrams with 
chargino exchange, which yield~\cite{KO}
\begin{equation}
|(\delta^u_{LL})_{22} - (\delta^u_{LL})_{11}| < 0.3 \times {m_{1/2} \over 
200~{\rm GeV}},
\label{KKbarW}
\end{equation}
where $(\delta^u_{LL})_{ii} \equiv (m^2_0)_{ii} / 
m_{\tilde q}^2$ is related to the difference between the second- and first- 
generation up-squark soft supersymmetry-breaking masses squared, in the 
canonical CKM basis. Assuming that $m_{\tilde q}^2 \simeq 6 m_{1/2}^2$, we 
see that (\ref{KKbarW}) is satisfied, for $m_{1/2} = 200$~GeV, by 
$(m^2_0)_{22} - (m^2_0)_{11} \lappeq 7 \times 10^4$~GeV$^2$. This 
condition is clearly satisfied for a substantial range of $m_0$ 
compatible with our lower limit (\ref{lowermzero}), even for maximal 
non-universality between $(m^2_0)_{22}$ and $(m^2_0)_{11}$. One should 
also consider box diagrams with gluino exchange, which yield~\cite{MPR}
\begin{equation}
|(\delta^d)_{12}| < 0.003 
\label{KKbarg}
\end{equation}
for $m_{1/2} = 200$~GeV, where $(\delta^d)_{12} \equiv (m^2_0)_{12} / 
m_{\tilde q}^2$ parameterizes a possible off-diagonal term in the 
down-squark soft supersymmetry-breaking mass-squared matrix, again in the 
CKM basis. For $m_{1/2} 
= 200$~GeV and again assuming $m_{\tilde q}^2 \simeq 6 m_{1/2}^2$, the 
upper limit (\ref{KKbarg}) requires $(m_0^2)_{12} \lappeq 700$~GeV$^2$.
Like the chargino constraint, this gluino constraint may be satisfied in a 
fairly natural way by an off-diagonal entry that is not very much smaller 
than our lower limit (\ref{lowermzero}). A complete analysis of the FCNI 
issue goes beyond the scope of this paper, and we refer the reader 
to~\cite{MPR} for a review.
However, we do note that the experimental upper limit on $\mu \to e 
\gamma$ decay does require the degeneracy between the slepton species to
be rather complete~\cite{EN}.
For this reason, in particular, it is desirable to find an explanation why 
the (necessarily) non-zero soft supersymmetry-breaking scalar masses should 
be generation-independent.

In this connection, we re-examine whether the no-scale hypothesis that
$m_0 = 0$~\cite{noscale} is really excluded. What we have shown above is 
that $m_0 \ne 0$
{\it at the GUT scale}. In a complete quantum theory of gravity, such as
string theory, the GUT scale is typically somewhat smaller than the Planck
scale. One can therefore imagine a scenario in which $m_0 = 0$ {\it at the
Planck scale}, with GUT interactions then renormalizing this starting
value: $m_0 \ne 0$ at the GUT scale. As an example how this might work, we
consider the minimal supersymmetric $SU(5)$ GUT. In this case, the soft
supersymmetry-breaking scalar masses of the 
$\mathbf 10$ and ${\mathbf \overline 5}$ 
representations of $SU(5)$ are renormalized differently above
$m_{GUT}$. In the one-loop approximation:
\begin{eqnarray}
{\partial m_{\mathbf 10}^2 \over \partial t} & = & {1 \over 16 \pi^2}
\left[ - {144 \over 5} g_5^2 m_{1/2}^2 \right], \\
{\partial m_{\mathbf \overline 5}^2 \over \partial t} & = & {1 \over 16 
\pi^2}
\left[ - {96 \over 5} g_5^2 m_{1/2}^2 \right],
\label{SU5RG}
\end{eqnarray}
where $t \equiv {\rm ln} (\mu^2 / \mu_0^2)$, $g_5$ is the $SU(5)$ gauge 
coupling, and we have neglected renormalization by Yukawa couplings.

To estimate the order of magnitude of the generation-independent value of
$m_0$ that may in this way be generated at the GUT scale, we insert the
value $g_5^2 / 4 \pi \simeq 1/20$ into (\ref{SU5RG}), finding $\partial
m_0^2 / \partial t \sim 0.1 \times m_{1/2}^2$. A complete analysis of the
coupled set of GUT renormalization-group equations for the soft
supersymmetry-breaking parameters goes beyond the scope of this paper, but
this rate of renormalization is clearly sufficient to generate values of
$m_0$ compatible with our lower limit (\ref{lowermzero}), even if the
effective Planck scale is only an order of magnitude beyond the GUT scale.

We conclude that, at least within the the CMSSM framework we have
discussed, experimental evidence indicates for the first time that soft
supersymmetry-breaking scalar masses {\it must be non-zero}, at least {\it
at the GUT scale}. This follows in general from the LEP lower limit on
$m_h$~\cite{LEPH} and the requirement that the LSP not be
charged~\cite{EHNOS}. As we have indicated, the required magnitude of
$m_0$ is not necessarily a disaster for FCNI, at least in the quark
sector. Moreover, the lower limit on $m_0$ is still compatible with
vanishing scalar masses {\it at the Planck scale}, as suggested by
no-scale models~\cite{noscale}.

\noindent{ {\bf Acknowledgments} } \\
\noindent  
The work of D.V.N. was partially supported by DOE grant
DE-F-G03-95-ER-40917, and that of K.A.O. by DOE grant
DE--FG02--94ER--40823.

\end{document}